\documentclass[twocolumn,superscriptaddress]{revtex4-1}
\usepackage{graphicx}% Include figure files
\usepackage{dcolumn}% Align table columns on decimal point
\usepackage{bm}% bold math
\usepackage[utf8]{inputenc}
\usepackage[T1]{fontenc}
\usepackage{mathptmx}
\usepackage{etoolbox}
\usepackage[dvipsnames]{xcolor}
\usepackage{upgreek}

\makeatletter
\def\@email#1#2{%
 \endgroup
 \patchcmd{\titleblock@produce}
  {\frontmatter@RRAPformat}
  {\frontmatter@RRAPformat{\produce@RRAP{*#1\href{mailto:#2}{#2}}}\frontmatter@RRAPformat}
  {}{}
}%
\makeatother

\begin{document}

% Use the \preprint command to place your local institutional report number 
% on the title page in preprint mode.
% Multiple \preprint commands are allowed.
%\preprint{}   

\title{Reduction of thermal instability of soliton states in coupled Kerr-microresonators} %Title of paper

% repeat the \author .. \affiliation  etc. as needed
% \email, \thanks, \homepage, \altaffiliation all apply to the current author.
% Explanatory text should go in the []'s, 
% actual e-mail address or url should go in the {}'s for \email and \homepage.
% Please use the appropriate macro for the type of information

% \affiliation command applies to all authors since the last \affiliation command. 
% The \affiliation command should follow the other information.

\author{Brandon D. Stone}
    \email[corresponding authors: ]{bstone898@unm.edu, drakete@unm.edu}
 \affiliation{Department of Physics and Astronomy, University of New Mexico, Albuquerque, New Mexico, USA}
 \affiliation{Center of High Technology Materials, University of New Mexico, Albuquerque, New Mexico, USA}
\author{Lala Rukh}
 \affiliation{Center of High Technology Materials, University of New Mexico, Albuquerque, New Mexico, USA}
 \affiliation{Optical Science and Engineering, University of New Mexico, Albuquerque, New Mexico, USA}
\author{Gabriel M. Colación}
 \affiliation{Center of High Technology Materials, University of New Mexico, Albuquerque, New Mexico, USA}
 \affiliation{Optical Science and Engineering, University of New Mexico, Albuquerque, New Mexico, USA}
\author{Tara E. Drake}
 \affiliation{Department of Physics and Astronomy, University of New Mexico, Albuquerque, New Mexico, USA}
 \affiliation{Center of High Technology Materials, University of New Mexico, Albuquerque, New Mexico, USA}
 \affiliation{Optical Science and Engineering, University of New Mexico, Albuquerque, New Mexico, USA}
%\homepage[]{Your web page}
%\thanks{}

%\date{\today}

\begin{abstract}

Kerr-microresonator frequency combs in integrated photonics waveguides are promising technologies for next-generation positioning, navigation, and timing applications, with advantages that include platforms that are mass-producible and CMOS-compatible and spectra that are phase-coherent and octave-spanning. Fundamental thermal noise in the resonator material typically limits the timing and frequency stability of a microcomb. The small optical mode volume of the microresonators exaggerates this effect, as it both increases the magnitude and shortens the timescale of thermodynamic fluctuations. In this work, we investigate thermal instability in silicon nitride microring resonators as well as techniques for reducing their effects on the microcomb light. We characterize the time-dependent thermal response in silicon nitride microring resonators through experimental measurements and finite element method simulations. Through fast control of the pump laser frequency, we reduce thermal recoil due to heating. Finally, we demonstrate the utility of a coupled microresonator system with tunable mode interactions to stabilize a soliton pulse against thermal shifts.

\end{abstract}

%\pacs{}% insert suggested PACS numbers in braces on next line

\maketitle %\maketitle must follow title, authors, abstract and \pacs

% Body of paper goes here. Use proper sectioning commands. 
% References should be done using the \cite, \ref, and \label commands

\section{Introduction}

Time-dependent thermodynamical fluctuations are a universal characteristic of matter at finite temperature, and their presence often sets a fundamental limit on our ability to detect and measure physical phenomena \cite{Callen:51}. In homogeneous media, temperature fluctuations have variance 
\begin{equation}
    \langle \delta T^2 \rangle = \frac{k_BT^2}{\rho CV},
\end{equation}
where $T$ is the average temperature, $k_B$ is the Boltzmann constant, $\rho$ is the density, $C$ is the specific heat, and $V$ is the volume in question \cite{Landau1980,Drake2020}. Temperature-dependent material properties likewise exhibit fluctuations. Additionally, light and interferometric measurements are influenced by thermal fluctuations in the surrounding medium, but the coupling between material thermal noise and optical phase noise is often also dependent on system details. A good example is phase noise in LIGO interferometric measurements, which originates from a combination of internal thermal noise in the test masses (e.g. mirror substrates), thermal expansion of the optical mirror coatings, and thermorefraction of the coating material, and is weighted by the size, shape, and penetration of the laser beams on the mirrors \cite{Levin:12}. The complicated relationship between temperature and optical phase noise also provides a mechanism for designing a system geometry with reduced thermal response--for example, an ultra-low frequency drift, temperature-insensitive photonic resonator for which thermal expansion and thermorefraction have been balanced \cite{Zhang:24}.

In the case of a pulse of light traveling in a Kerr-microresonator, thermal noise in the resonator material will influence the properties of the light primarily through thermorefraction and thermal expansion \cite{Carmon:04}. As can be seen in equation 1, the small optical mode volume of these resonators increases the amplitude of the fluctuations of temperature-dependent optical properties. Additionally, the thermalization time in microresonators is short (on the order of 1 microsecond \cite{Carmon:04,Briles2020}), which presents challenges for suppressing thermally driven fluctuations via electronic feedback-based control loops. Finally, the tight confinement of light needed for many microresonator applications often increases the material temperature via optical absorption, which both enhances fluctuations and makes cryogenic cooling schemes difficult.

In this work, we explore the effects of thermodynamics and thermally driven instability in silicon nitride microring resonators. The thermalization time of our microresonators is investigated experimentally and with finite element simulations. We then implement control over resonator heating during thermalization by utilizing rapid changes of the pump laser frequency. Finally, we investigate the use of an auxiliary microresonator with tunable coupling to the main microresonator for stabilizing a soliton pulse against resonator thermal shifts.

\section{Experimental Design}

\begin{figure}
    \centering
    \includegraphics[width=1\linewidth]{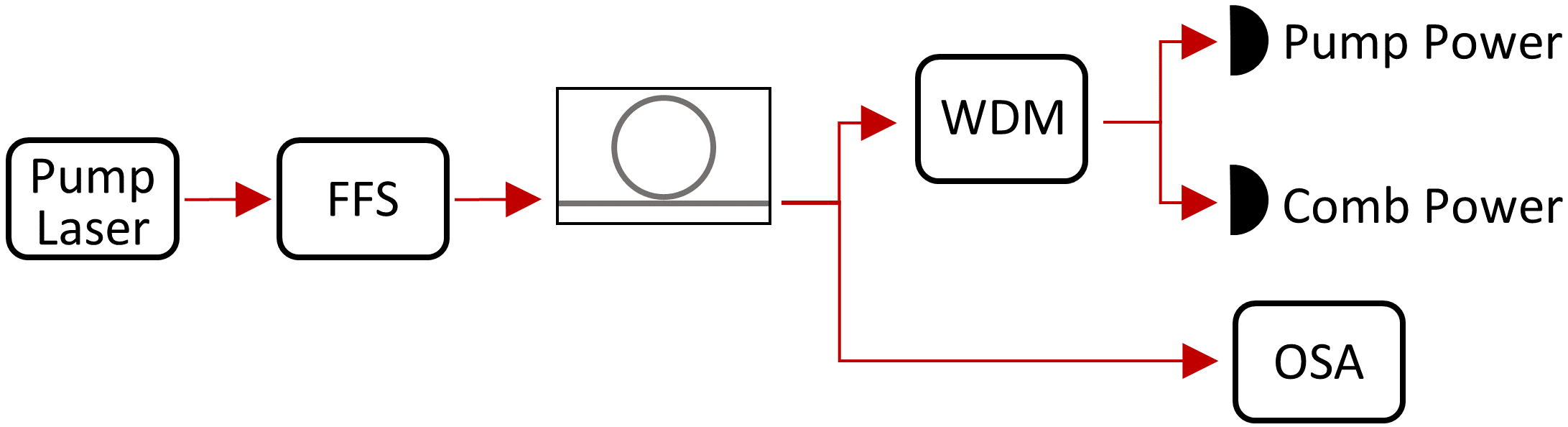}
    \caption{Diagram of the experimental setup for soliton generation. A pump laser is sent through our fast frequency sweeper (FFS) system and then into an erbium doped fiber amplifier (not shown). The optical waveform in the resonator is analyzed using a wavelength-division multiplexer (WDM) and two photodetectors, as well as an optical spectrum analyzer (OSA).}
    \label{fig1}
\end{figure}

Figure 1 illustrates our experimental design for generating solitons in silicon nitride (Si$_3$N$_4$) microresonators. To support generation of a dissipative Kerr soliton (DKS), we use a Si$_3$N$_4$ waveguide ring resonator with dimensions chosen for anomalous dispersion at telecom C-band wavelengths. The single-frequency (CW) pump laser is an external cavity diode laser (ECDL) with enhanced optical frequency control using a single-sideband electro-optic modulator, following references \cite{Briles2020}, \cite{Stone:18}, and \cite{Wang:15}. In this fast frequency sweeper (FFS) setup, a high slew-rate voltage controlled oscillator (VCO) controls the wavelength of the modulator sideband used as the CW pump, and all other sidebands and the carrier are suppressed. By sending a custom linear voltage sweep to the VCO, we are able to change the pump laser frequency by 6 GHz at speeds up to 60 GHz/$\upmu$s. The sideband/pump is coupled into the microring to generate a DKS pulse. The soliton characteristics are investigated using an optical spectrum analyzer and photodetectors on the pass and reflect arms of a C-band wavelength division multiplexer (WDM). The WDM is chosen to pass light in a narrow window around the pump laser frequency, allowing us to separately measure the pump and comb power.

\section{Results}

\begin{figure}[h!]
    \centering
    \includegraphics[width=1\linewidth]{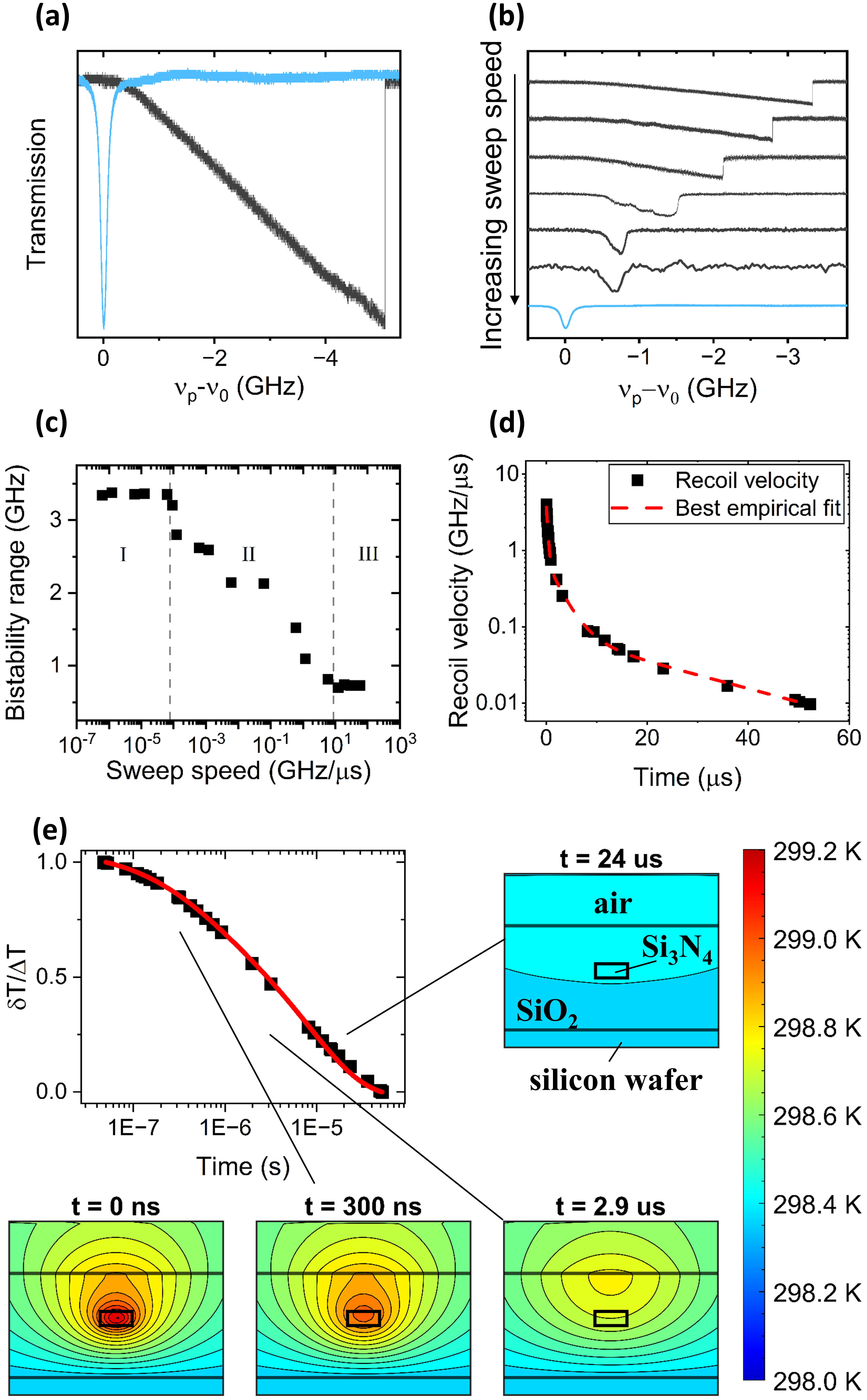}
    \caption{Magnitude and timescale of thermal response. (a) Two adiabatic laser frequency sweeps through a resonance, at high optical power (black) one at low optical power (blue). (b) Distortion of resonance at high power with laser frequency swept from blue to red of the resonance at varying speeds. The blue trace is taken at low power for reference (using an adiabatic sweep). (c) Total resonance shift as a function of laser frequency sweep speed. We identify three distinct regions based on the mode response and indicated amount of heating. (d) Measured thermalization of the microring after a sharp drop in intracavity power, using the frequency of a probed resonance. An empirical fit to the sum of three exponentials with different decay rates is shown in red. (e) Comparison of cooling the probed mode in the experiment (black) and cooling of the silicon nitride microring using a COMSOL simulation (red). Temperature contour plots of the resonator cross section are taken at hot equilibrium (t = 0 ns), and then at each decay constant extracted from the empirical fit (t = 300 ns, 2.9 $\upmu$s, and 24 $\upmu$s).}
    \label{fig2}
\end{figure}

To mitigate thermal effects in our system, we first characterize the microresonator's thermal response. In these resonators, the thermorefractive effect, $\frac{dn}{dT}>0$, leads to an increase of optical path length as the resonator heats from absorption. When performing a wavelength sweep across the resonance at low optical power, heating is negligible and we observe a Lorentzian lineshape with a width set by the resonator quality factor (blue trace in figure 2a). At higher optical powers, the system exhibits detuning-dependent absorptive heating, dynamically red-shifting the resonance towards lower frequencies and distorting and broadening its optical lineshape, as described in \cite{Carmon:04} (black trace in figure 2a). The magnitude of this shift can be reduced if the laser frequency is swept faster than the resonator can fully thermalize with the coupled laser light. In figure 2b, we observe the change of a resonance lineshape as the optical power of the pump laser is kept constant but the rate of laser frequency change is increased. The total resonance shift (the bistability range, see reference \cite{Carmon:04}) as a function of the rate of change of the pump laser frequency (sweep speed) is shown in figure 2c. We observe 3 regions in this data. Region I describes “adiabatic” sweeps, where the sweep speed is slow enough that the resonator is in thermal equilibrium throughout the entire sweep. In region II, we see a reduction in the thermally induced shift as a function of the sweep speed, indicating that the time spent coupling light into in the resonator is on the order of the resonator's thermalization time. In region III, the pump laser frequency is changed fast enough that absorptive heating is negligible, and the mode is no longer distorted. Even at the fastest sweep speeds, the mode is red-shifted from its low power frequency (figure 2b, blue trace) by the (near-instantaneous) Kerr effect. We measure a minimum resonance shift of 730(70) MHz, which is close to our calculated Kerr shift of 650 MHz for 33 mW in the waveguide.

We use a pump-probe measurement to directly measure the time-dependent thermalization of the microresonator (following reference \cite{Brasch:16}). The pump laser is swept adiabatically through an optical resonance, while a low-power probe laser is kept at constant frequency slightly blue-detuned from a different resonance. When the pump reaches the end of the bistability range, its power in the microring drops to zero, and the resonator rapidly cools. 
During cooling, the second resonance will sweep across the constant-frequency probe laser. By measuring the duration of resonance-probe coupling and comparing it to the resonance's natural linewidth, we calculate the microresonator's rate of thermal relaxation, or recoil velocity. By changing the probe laser's initial detuning from the second resonance, we sample the relaxation velocity for the first 50 microseconds of cooling (figure 2d). Although we do not necessarily expect the thermal response of the resonator to have a closed-form analytic solution \cite{Huang:19}, following thermal mode decomposition \cite{Gorodetsky:04,Matsko:07,Kondratiev:18,Pavlov:23}, we empirically fit the change in recoil velocity to the sum of three exponential decays. We find decay time constants of 300(20) ns, 2.9(2) $\upmu$s, and 24(1) $\upmu$s.

We further investigate thermal recoil in our microresonator with a finite element method (COMSOL) simulation. (Simulation details are found in Supplementary Material.) The resonator is heated via optical absorption and allowed to come to a warm equilibrium, and then the heat source is turned off. We track the average temperature of the waveguide and compare the fractional temperature change to that calculated from the experimentally measured recoil velocity (figure 2e, main panel). We also provide a more qualitative understanding via four temperature contour plots showing snapshots of a cross section of the silicon nitride microring, silica cladding, and silicon/air environment. The snapshots, taken directly after the heat source is turned off (t = 0 ns) and at the timescales found through empirical fitting, highlight the thermalization of separate domains in the devices, which happen at different rates. Using the cross-sectional temperature maps, we interpret our thermal recoil data to reflect an initial fast thermalization of the optical mode volume within the silicon nitride waveguide to its surroundings (t = 300 ns), a slightly slower thermalization related to the silica cladding around (and particularly above) the Si$_3$N$_4$ waveguide (t = 2.9 $\upmu$s), and a gradual thermalization of the entire device with the environment (t = 24 $\upmu$s).

\begin{figure}
    \centering
    \includegraphics[width=1\linewidth]{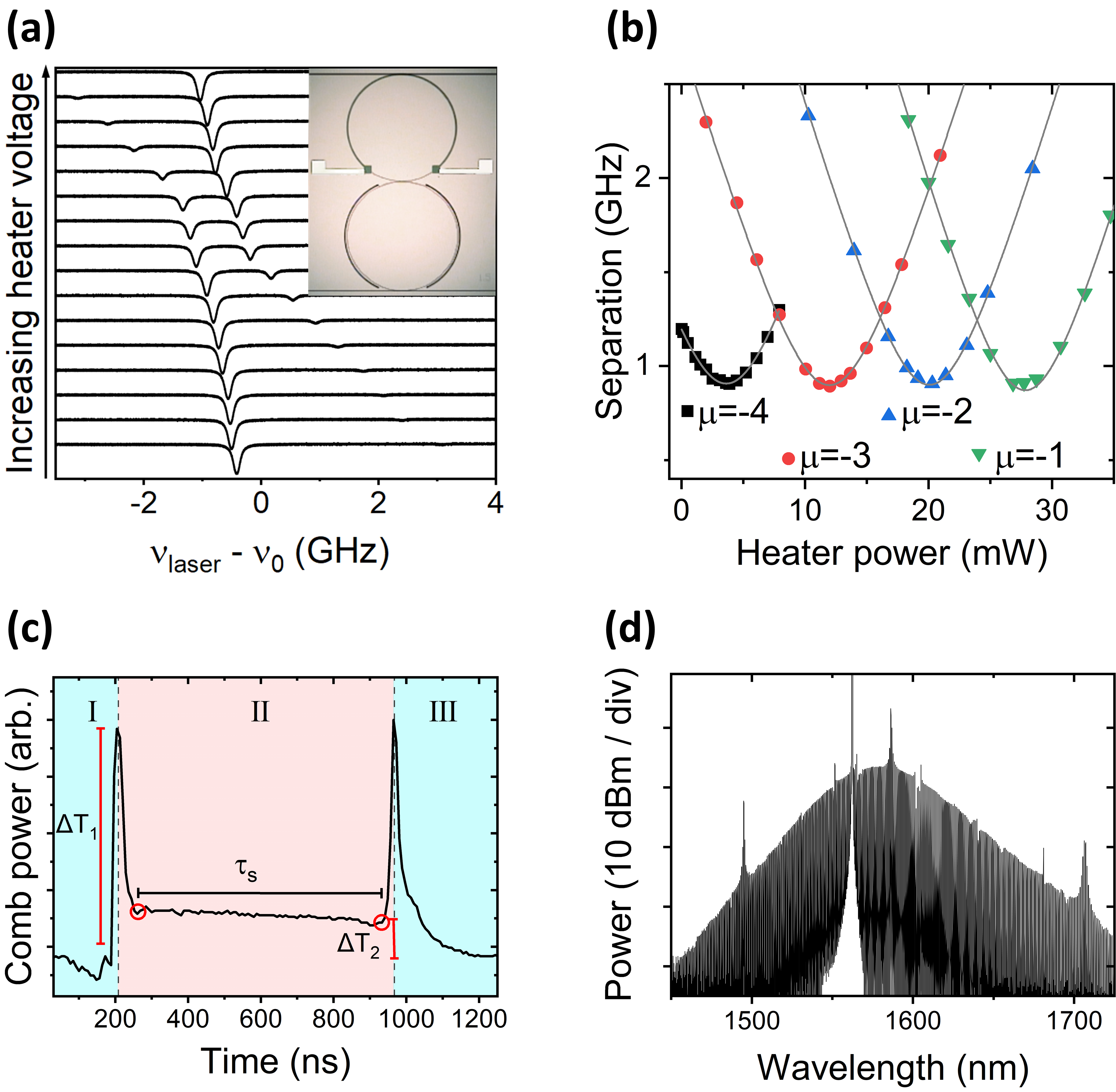}
    \caption{Coupled resonator details. (a) Avoided mode crossing of two modes in the main and auxiliary resonators, as seen in pump laser transmission. The inset is a micrograph of the coupled Si$_3$N$_4$ microring resonators. The main resonator is the bottom ring, and the top ring is equipped with a resistive heater. (b) Measured separation of four main-auxiliary mode pairs nearest to and red of the pumped mode as a function of power dissipated in the heater. (c) A short-lived soliton generated using fast laser sweeping. The soliton lifetime ($\tau_s$) is shown. (d) Spectrum of a long-lived soliton, achieved with optimization of laser sweep parameters.}
    \label{fig3}
\end{figure}

In the second half of this paper, we utilize an auxiliary microring resonator evanescently coupled to the main microring resonator (from figure 2) in order to investigate the use of tunable mode crossings to increase the stability and lifetimes of DKS states. Our work borrows from and builds on previous work developing coupled-mode formalism \cite{Haus:91,Savchenkov:12} and on experiments utilizing coupled resonators for increasing microcomb efficiency \cite{Miller:15,Helgason:23}, for altering resonator dispersion \cite{Xue:15,Fujii:18,Kim:19,Helgason:21,Triscari:23}, and for the realization of novel optical pulse states \cite{Rowley:22,Yuan:23}.

The auxiliary resonator in our system has a 2\% increase in free spectral range (FSR) compared to the main resonator and is equipped with a resistive heater, allowing us to increase the optical path length via the thermorefractive effect. This allows us to control the interactions between optical modes of the resonators. These mode interactions are described by:
\begin{equation}
\nu^{(s,as)} = \frac{\nu_m^{main} + \nu_n^{aux}}{2} \pm \sqrt{\frac{(\nu_m^{main}-\nu_n^{aux})^2}{4}+\frac{\kappa^2}{4}},
\end{equation}
where $\nu^{(s,as)}$ are the resonant frequencies of the optical modes when dressed by the interaction, $\nu_m^{main}$ and $\nu_n^{aux}$ are the bare resonance frequencies, and $\kappa$ is the coupling rate parameterizing the strength of modal interaction \cite{Xue:15}. By driving current through the heater, we induce a linear shift in $\nu_n^{aux}$. In our system, heating the auxiliary resonator to bring an auxiliary-main mode pair into resonance will typically only affect that pair, as the next-nearest pairs will be detuned farther than a typical coupling rate. 

In figure 3, we present the details of the coupled resonators (figure 3a inset) and the mode interactions and set up the measurement of DKS lifetime as an indication of the resilience of the optical state to thermal fluctuations. Figure 3a shows the transmission of a low intensity laser coupled into the main resonator as an auxiliary resonance is tuned across a main resonance. The plot shows an avoided crossing as expected from equation 2, with the smallest separation of the dressed modes equal to the coupling rate term. Figure 3b characterizes the avoided crossings for the four consecutive mode pairs red of the pumped mode (modes $\upmu$ = -1 to -4), which are also the first four mode crossings accessed by heating. The coupling rate is measured to be approximately 0.9 GHz for all of these mode pairs, and the separation between the dressed resonances, $\Delta \nu$, closely fits the expected form
\begin{equation}
    \Delta \nu = \nu^{(s)} - \nu^{(as)} = \sqrt{(\nu_m^{main}-\nu_n^{aux})^2 + {\kappa^2}},
\end{equation}
also plotted (gray lines).

\begin{figure*}
    \centering
    \includegraphics[width=.9\linewidth]{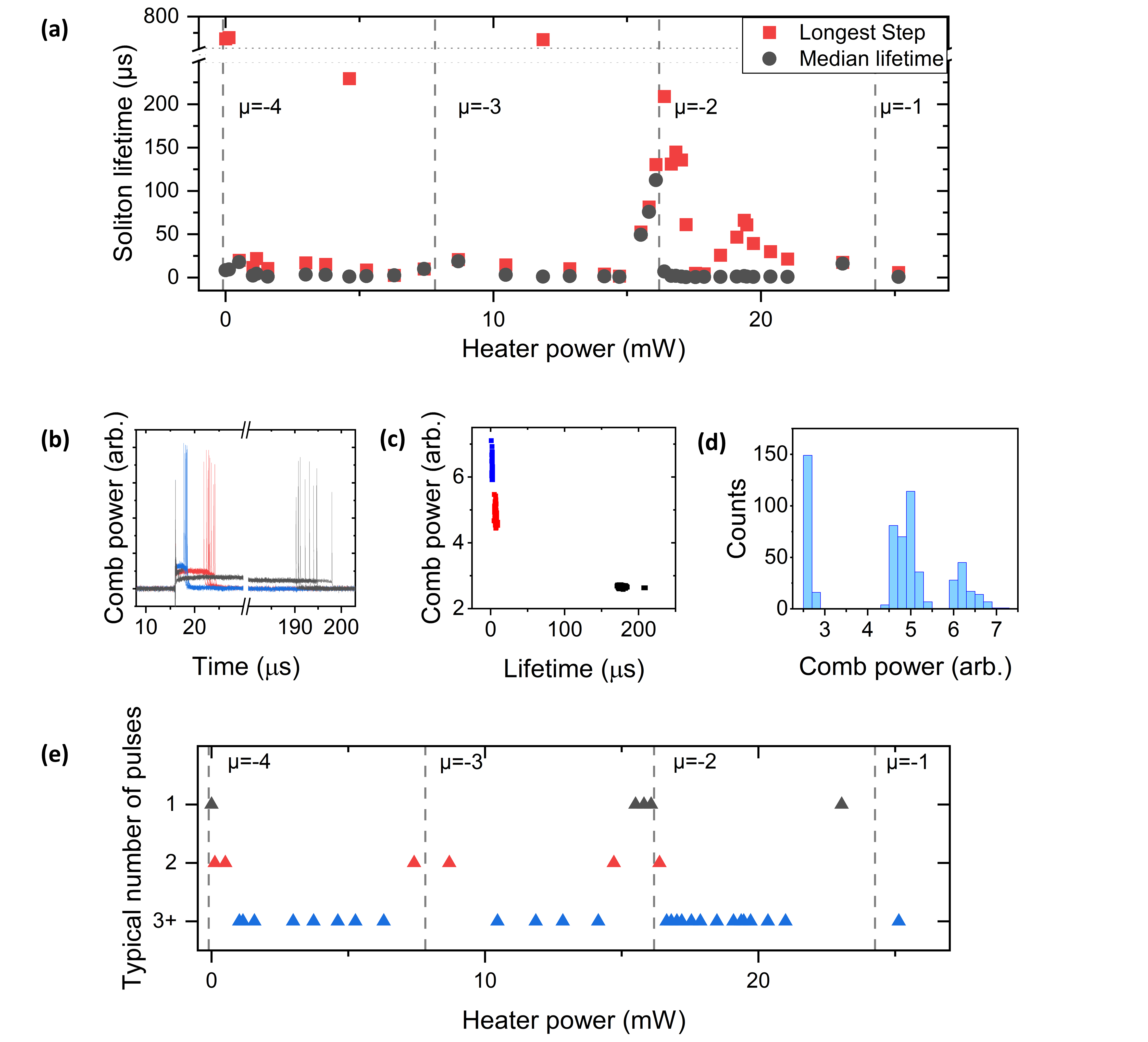}
    \caption{Effect of mode interactions on soliton lifetime. (a) Scatter plot of soliton lifetime as a function of power dissipated by the resistive heater, showing the median and longest lifetime reported out of 660 sweeps for each data point. (b) Comb power overlaid from multiple sweeps from a single data point. (c) A scatter plot of the average comb power and state lifetime. (d) Histogram of comb powers from the same data set, showing 3 distinct groupings. (e) Most commonly observed number of soliton pulses in the cavity (omitting sweeps with zero solitons).}
    \label{enter:label}
\end{figure*}

By using the fast frequency sweeper technique described above, we demonstrate soliton generation in these devices. With an adiabatic sweep, the natural lifetime of the soliton is set by the thermal recoil of the resonator and is typically < 100 ns in our devices. When using a pump laser frequency sweep on the order of the resonator thermalization time, the soliton state lives longer. Figure 3c shows a soliton step in the comb power using this technique, which we characterize as having three distinct regions (following reference \cite{Briles2020}). In region I, the frequency sweep is in progress, with the pump laser blue-detuned from resonance. We observe a sharp increase in comb power as we sweep closer to the mode and four-wave mixing begins. During this time, the resonator heats by an amount $\Delta T_1$, roughly dependent on the optical power in the resonator weighted by the time-dependent thermalization and thus influenced by the pump laser sweep speed. In region II, the soliton pulse has formed as the laser becomes red-detuned. The DKS state has less optical comb power than its precursor, and the abrupt change in total resonator power can cause thermal recoil. The comb power associated with the DKS state has an equilibrium temperature above the cold resonator temperature of $\Delta T_2$. If $\Delta T_1 << \Delta T_2$, then the resonator heats, causing a reduction in the pump-resonance detuning until the laser passes back through the mode. This results in a transition into region III, where the laser is not sweeping, but is blue-detuned from resonance. The soliton lifetime, $\tau_s$ in figure 3c, is defined as the duration of the soliton state, as inferred from the comb power level. If the pump laser sweep is carefully chosen such that $\Delta T_1 - \Delta T_2$ is small enough that the subsequent heating or cooling after landing in a soliton state doesn't create a significant change in detuning, the resonator will reach a new thermal equilibrium with a stable soliton in the resonator (shown in figure 3d).

Using the definition of soliton lifetime from figure 3c, we investigate the effect of avoided mode crossings on soliton lifetime. Specifically, we want to see whether the presence of a nearby mode crossing has an effect on the soliton lifetime. We use the fast sweeping technique described, choosing a sweep speed that enhances the natural lifetime but it is not quite fast enough to fully stabilize the soliton; this deliberate choice allows us to more easily measure the effect of the auxiliary resonance. 
In this way, the soliton lifetime is used as a metric of the optical state's resilience to thermal shifts under varying interaction strength with the auxiliary resonator (controlled via dissipative power in the resistive heater).

Figure 4a shows the median (black) and longest recorded (red) soliton lifetimes for 660 laser sweeps per data point. The dashed vertical lines indicate the heater powers corresponding to the closest approach of the respective labeled mode. These are shifted by 950 MHz from the crossings characterized at low pump power in 3b due to the heating from the pump laser. We observe that the presence of an auxiliary mode does influence the soliton lifetime. This is most obvious at $\upmu=-2$, but there are small increases in median lifetime near $\upmu=-3$ and $-4$ as well. The stronger effect at $\upmu=-2$ is probably related to its closer proximity to the pumped resonance. We do not observe this behavior at the $\upmu=-1$ main-aux. crossing, because at this heater power, we find that an auxiliary resonance is close enough to the pumped resonance to inhibit soliton state formation by coupling pump laser power into the auxiliary resonator. It is possible that the increase in longest soliton lifetime we observe between $\upmu=-2$ and $\upmu=-1$ is related to the expected increased stability near $\upmu=-1$, but our results are inconclusive.

Figures 4(b-d) provide further detail on soliton lifetimes at a single auxiliary resonator temperature in figure 4a (heater power = 16.39 mW, slightly below the $\upmu=-2$ crossing). 
In figure 4b, we plot a selection of solitons observed at this heater power. We identify three state types, distinguishable by both comb power in the soliton (i.e. in region II from figure 3c) and by state lifetime. These groupings are even more obvious in figures 4c and 4d. We interpret these groups as states with varying numbers of solitons in the cavity. In figures 4b and 4c, the black states are the lowest power and identified at one soliton, the red are two solitons, and the blue are a three-soliton state. (The long-lived soliton spectrum in figure 3d corresponds to a comb power level equal to the black states, albeit using a faster laser sweep speed.) We also observe distinct lifetimes for each state type. The single soliton state is the longest-lived (likely due to our optimization), and multi-solitons have significantly shorter lifetimes, as their optical power is not as well matched to the initial heating during the laser frequency sweep. Similarly, we observe that the presence of an auxiliary resonance affects the relative distribution of soliton states (figure 4e). Near the crossings, soliton states are more likely to have only one pulse, while farther away the soliton states tend to have multiple pulses with shorter lifetimes. The rich dynamics of optical state formation near an auxiliary mode crossing is further explored in the Supplementary Material.

In conclusion, we have thoroughly investigated the thermal dynamics associated with soliton state formation in Kerr-microring resonators. We benchmark thermalization of the microresonator in several ways. To begin with, we use a fast laser frequency sweep to achieve variable dwell times in the resonance and reduce the thermal bistability. We also measure the temporal change of an optical resonance after an abrupt decoupling of laser light from the cavity. Finally, we simulate abrupt cooling of the resonator using the finite element method (COMSOL). We then investigate a scheme for inhibiting optical state instability in the face of thermal shifts through the use of a secondary resonator with a nearly-degenerate mode red-detuned from the pumped mode. We observe that a mode degeneracy can stabilize a dissipative Kerr soliton (DKS) against thermal recoil, thereby increasing its lifetime. We also observe a connection between stabilizing mode crossings and our ability to deterministically create a DKS state with a desired number of pulses.

We believe that the use of auxiliary resonators to reliably generate optical states and reduce noise associated with material thermodynamics is a broadly applicable technique with many potential implications for the achievement of low-noise microcombs.

\section{Methods}
All experimental data in this paper were taken using dual coupled microring resonators made of stoichiometric silicon nitride. The rings have radii of 260 $\upmu$m and 254.8 $\upmu$m, and the cross-sectional size of the waveguide is 1.5 $\upmu$m wide by 800 nm tall. The two rings are evanescently coupled with a gap of 600 nm and a 400 nm coupling gap between each ring and the bus waveguides. The resistive heaters on the auxiliary rings are controlled using DC contact probes connected to a voltage supply. The heaters themselves provide 1.94 k$\Omega$ resistance, allowing them to dissipate power to the ring. The resonators were fabricated at Advanced Micro Foundry Pte Ltd.

\section{Supplementary Material}
Further information regarding the limitations of the FFS, the measurement of thermal recoil velocity, the finite element method simulation, and additional soliton lifetime statistics can be found in the supplementary material document.

\section{Acknowledgments}
We gratefully acknowledge financial support from AFOSR (FA9550-22-1-0174), NSF EPSCoR (2217786), DARPA GRYPHON (HR001122C0017), NSF (2340973), and UNM Women in STEM awards. We thank Professor Chee Wei Wong and his team at UCLA for the use of the coupled resonator devices.

% Create the reference section using BibTeX:
%\section{References}
\bibliography{coupledresbib}

\end{document}

\title{Supplementary Material for ``Reduction of thermal instability of soliton states in coupled Kerr-microresonators"}
%\date{\today}
\author{Brandon D. Stone}
\affiliation{Department of Physics and Astronomy, University of New Mexico, Albuquerque, New Mexico, USA}
\affiliation{Center of High Technology Materials, University of New Mexico, Albuquerque, New Mexico, USA}
\author{Lala Rukh}
\affiliation{Center of High Technology Materials, University of New Mexico, Albuquerque, New Mexico, USA}
\affiliation{Optical Science and Engineering, University of New Mexico, Albuquerque, New Mexico, USA}
\author{Gabriel M. Colación}
\affiliation{Center of High Technology Materials, University of New Mexico, Albuquerque, New Mexico, USA}
\affiliation{Optical Science and Engineering, University of New Mexico, Albuquerque, New Mexico, USA}
\author{Tara E. Drake}
\affiliation{Department of Physics and Astronomy, University of New Mexico, Albuquerque, New Mexico, USA}
\affiliation{Center of High Technology Materials, University of New Mexico, Albuquerque, New Mexico, USA}
\affiliation{Optical Science and Engineering, University of New Mexico, Albuquerque, New Mexico, USA}
%\pacs{}

\maketitle

\section{Limitations of the Fast Frequency Sweeper system}

%\textcolor{red}{- Maybe add names for specific components? \\
%- Tara: check to see how much of this setup is clear in TCB and JS works \\
%}

In order to characterize the performance of the fast frequency sweeper (FFS), we send the modulated sideband through a Mach-Zehnder interferometer (MZI), which monitors the relative pump laser frequency change, and we compare the desired (programmed) sweep rate to the measured sweep rate over many orders of magnitude of sweep speed. The MZI fringe spacing is calibrated 194(2) MHz using an absolute wavemeter measurement. By finding the time between each peak, we have a measurement of the sweep rate throughout the sweep. Figure \ref{fig:FFS}a shows an MZI trace for a sweep with a duration of 100 ms over 6 GHz of optical frequency tuning. 

\renewcommand{\thefigure}{S\arabic{figure}}
\renewcommand{\thetable}{S\arabic{table}}
\renewcommand{\theequation}{S\arabic{equation}}
\begin{figure}[h!]
    \centering
    \includegraphics[width=1\linewidth]{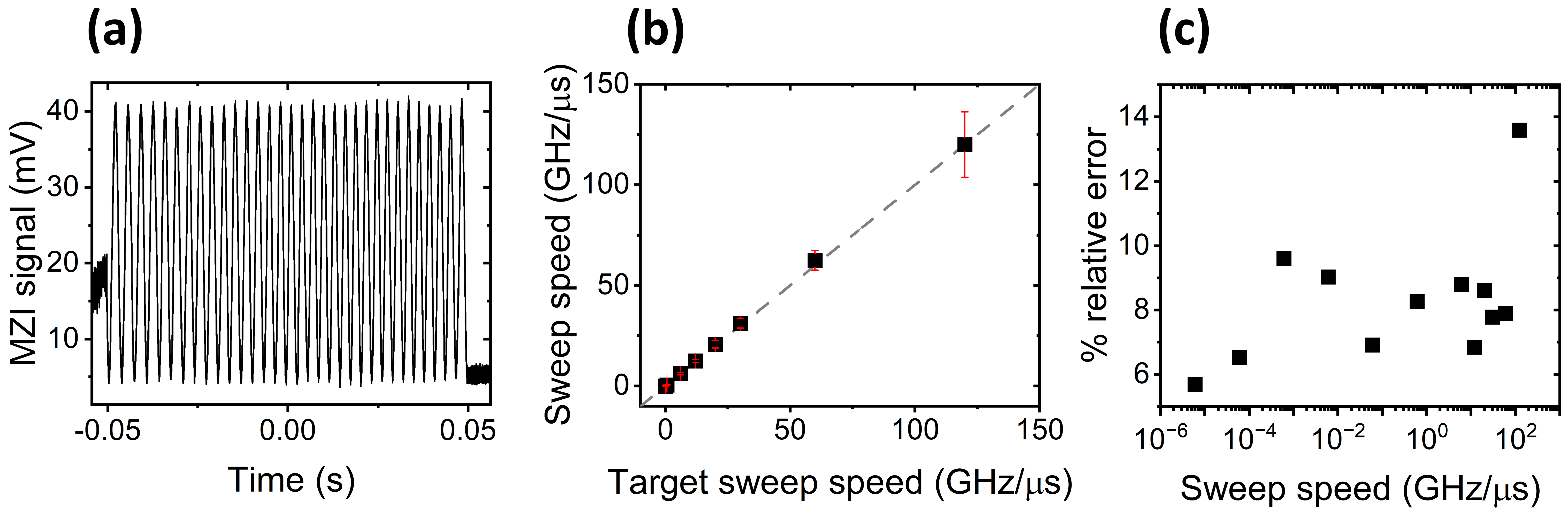}
    \caption{Fast Frequency Sweeper system performance. (a) Mach-Zehnder interferometer signal calibrating the single sideband frequency change for a sweep of 6 GHz over a 100 ms duration. (b) Average sweep speed measured using MZI signal for the full range of sweep rates used in the experiment. The dashed gray line is y=x, signifying a measured sweep rate that matches the programmed sweep rate. The error bars show the standard deviation of sweep speeds (measured fringe-to-fringe) during a single sweep. (c) Relative statistical (standard) deviation as a function of sweep speed.}
    \label{fig:FFS}
\end{figure}

There are several potential limiting system components that might lead to deviations between desired pump frequency sweep speed and the actual speed. We expect the slew rate of the voltage controlled oscillator (VCO) providing the RF signal to the single-sideband modulator to provide one such limitation; this component has a manufacturer specified rise time of 100 ns. A 100 ns sweep over the 6 GHz accessible range of the VCO gives a sweep rate of 60 GHz/$\upmu$s. In figure \ref{fig:FFS}b, we show that the programmed and measured average sweep speeds are in very good agreement within the range of 6 kHz/$\upmu$s to 120 GHz/$\upmu$s. For t $\leq$ 100 ns, we observe that the relative sweep rate variation across the sweep remains less than 10\% (figure \ref{fig:FFS}c). The fastest sweep with duration 50 ns (120 GHz/$\upmu$s) shows a less constant rate across the entire sweep, which is not unexpected given VCO specifications. Other potential limiting components might be the erbium-doped fiber laser (EDFA), which amplifies the swept sideband, the RF amplifier after the VCO, and/or the programmed voltage sweep that controls the VCO, but as our measurements in figure \ref{fig:FFS} include all these components, we do not expect them to contribute for sweep rate at 60 GHz/$\upmu$s or slower.

\section{Thermal recoil velocity measurement}

\begin{figure} [h!]
    \centering
    \includegraphics[width=0.9\linewidth]{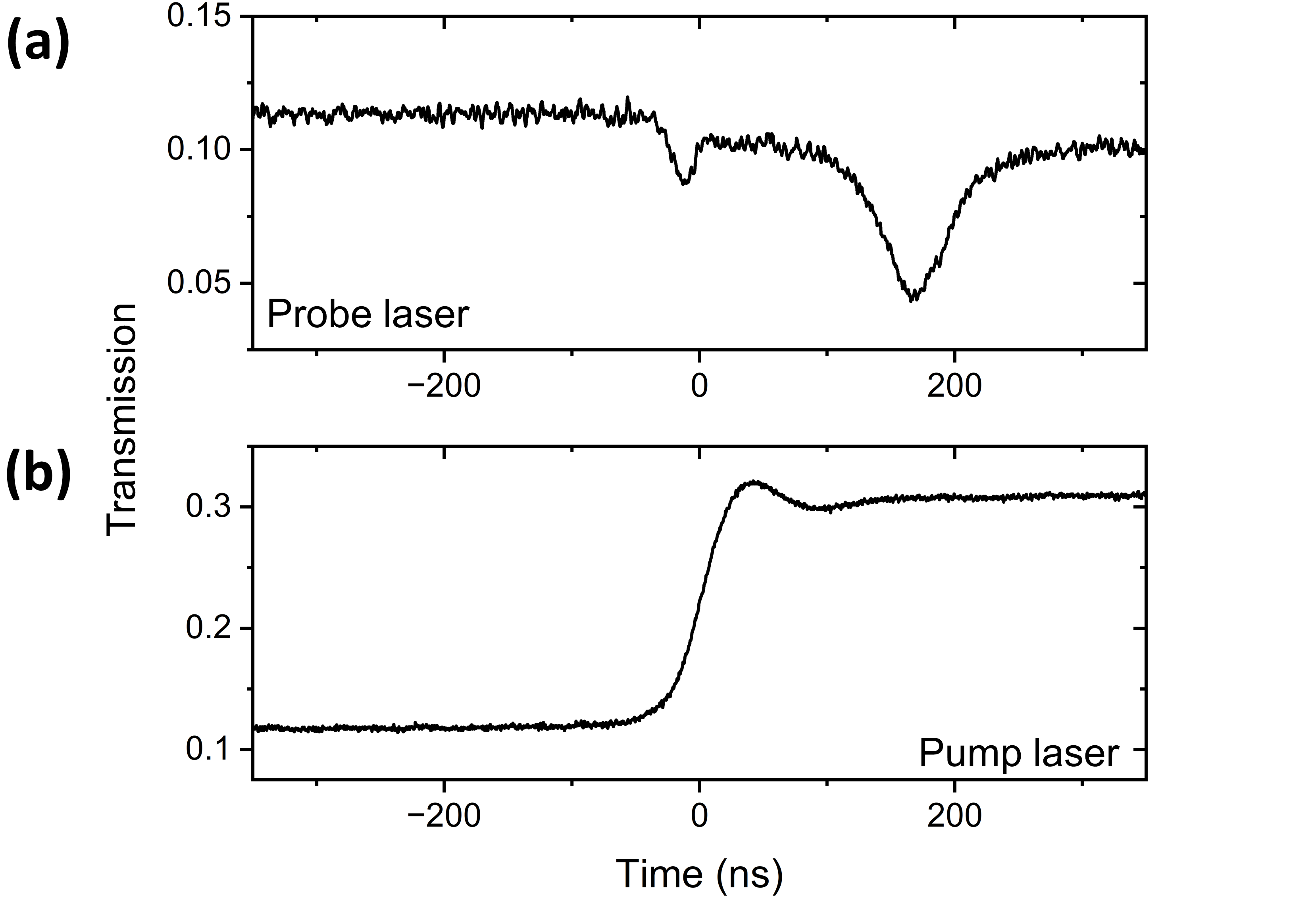}
    \caption{Thermal relaxation velocity measurement. (a) Counter-propagating probe transmission signal. The probe laser is kept at constant frequency on the blue side of a resonance. When the resonator starts to cool at the end of the thermal bistability region, we see the resonance scan through the probe laser. (b) Forward-propagating pump laser transmission signal. This is taken during an adiabatically slow sweep (i.e. maintaining thermal equilibrium throughout). The abrupt increase in transmission marks the end of the bistability range. (The sharpness of the rising edge is likely limited by photodetector bandwidth.)}
    \label{fig:pump-probe}
\end{figure}

To measure the thermal recoil velocity of the resonator directly, we utilize a pump-probe setup. \cite{Brasch:16} The experimental setup seen in figure 1 in the main text is modified to include circulators before and after the microresonator. This allows us to use a backward-propagating, low power, tunable CW laser as a probe. A tunable filter and WDM were needed to isolate the probe light from pump laser back-reflections in the resonator (significant due to the larger power of the pump laser). The forward- and backward-propagating laser transmissions are seen in figure \ref{fig:pump-probe} for a given probe laser-probe resonance detuning. The feature seen in the probe transmission as we fall out of the thermal triangle is chosen as $t$ = 0, as it marks the beginning of thermal recoil with the pump laser is abruptly shifted off-resonance. The probed resonance is then fit to a Lorenztian lineshape. The lineshape center tells us the time that we are sampling, and the width of the resonance is directly related to the rate of resonator cooling (i.e. thermal relaxation/recoil). Under the assumption that the cooling rate is constant for our measurement (accurate when the mode in figure \ref{fig:pump-probe}a is symmetric), we determine the recoil velocity as $ \dot\nu=\frac{\Delta\nu}{\Delta t}$, where $\Delta\nu$ is the Lorenztian linewidth in MHz measured separately and $\Delta t$ is the linewidth in time (i.e. the time it takes for the resonance to change frequency by one linewidth). This method is effective for measuring the thermalization rates close to $t$ = 0, but at large $t$ we observe asymmetric lineshapes, indicating cooling rates that are too slow to measure in this way. However, fast timescales dominate the dynamics of soliton formation in which we are interested, and we are able to measure thermal recoil velocities up to t = 52 $\upmu$s after a laser power change.

In figure 2e in the main text, these results have been converted to a fractional change in temperature occurring in the measurement window. To do this, we numerically integrate the relaxation velocity data to obtain a measurement of the accumulated frequency shift at each point in time. Then, we use the relation \cite{Carmon:04}
\begin{equation}
    \Delta T = \frac{\lambda_r - \lambda_0}{\frac{\frac{dn}{dT}}{n_0}\lambda_0}
\end{equation}
to determine the magnitude of the temperature shift at each data point. We normalize this data for comparison with the temperatures found using COMSOL simulations.

\section{Finite element method simulation details} 

\begin{figure}
    \centering
    \includegraphics[width=0.85\linewidth]{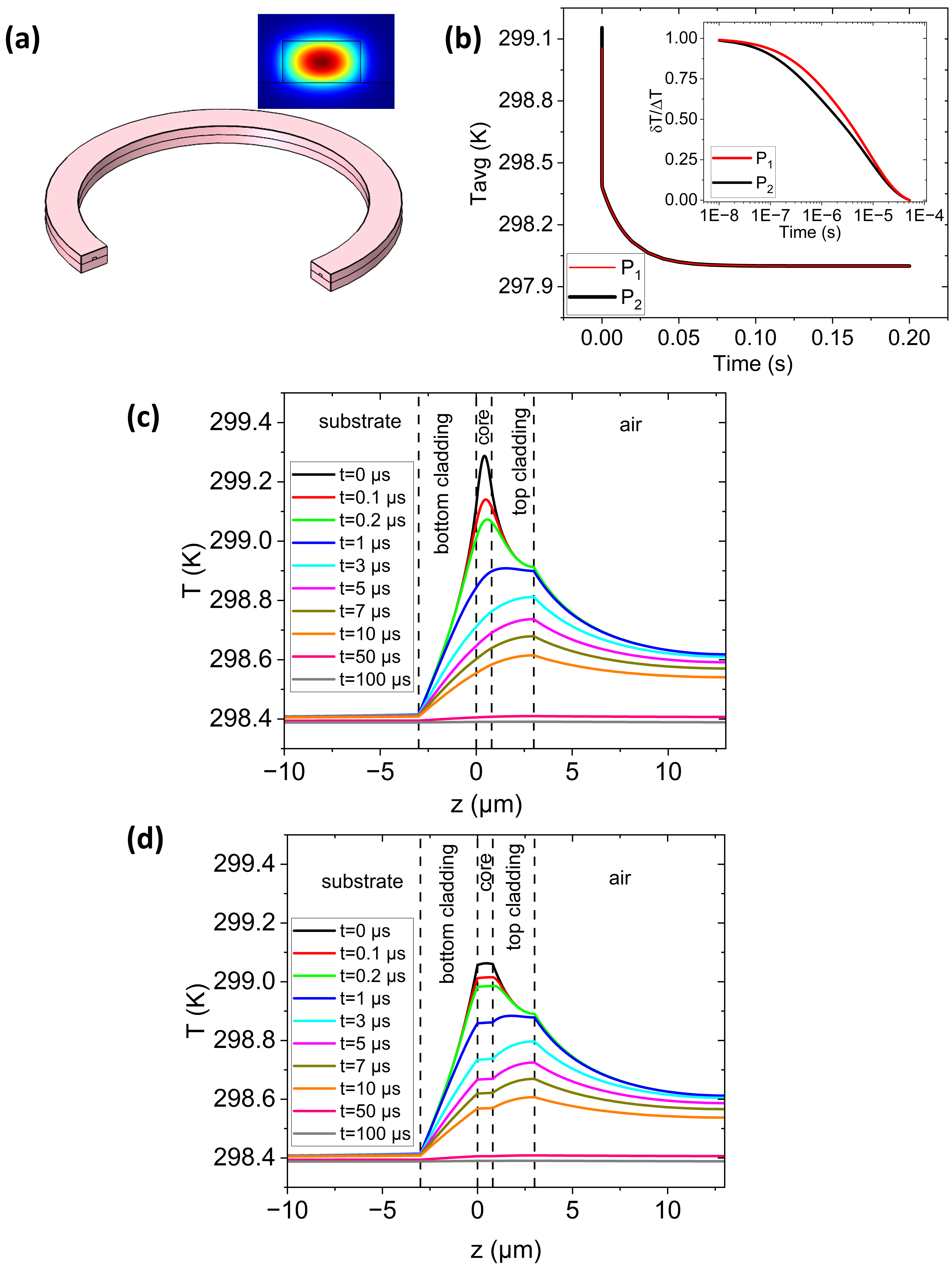}
    \caption{FEM simulation of thermal response. (a) 3D model of a microring resonator. Inset: TE fundamental mode used as the distributed heat source. (b) Transient response of the average temperature of the silicon nitride for parameter sets P$_1$ (red) and P$_2$ (black). Inset: fractional change in silicon nitride average temperature for the first 50 $\upmu$s of cooling, using both sets of parameters. (c) 1D temperature profile along the z-axis at various snapshots in time for property set P$_2$. (d) 1D temperature profiles along the z-axis for property set P$_1$.}
    \label{fig:sim}
\end{figure}

Finite element method (FEM) COMSOL simulations are used to model the transient thermal response of the silicon nitride (Si$_{3}$N$_{4}$) ring resonator. The resonator is modeled using axial symmetry with a Si$_{3}$N$_{4}$ core of width 1.5 $\upmu$m and height 0.8 $\upmu$m surrounded by 6 $\upmu$m thick silica cladding, with silicon as the lower substrate and air above the silica top cladding. Figure \ref{fig:sim}a shows the 3D geometry of the ring resonator. This simulation follows the method outlined in references \citenum{kondratiev_thermorefractive_2018} and \citenum{huang_thermorefractive_2019}, using the COMSOL Wave Optics and Heat Transfer in Solids modules. We use the Electromagnetic Wave, Frequency Domain (ewfd) physics interface to find the profile of a fundamental transverse electric (TE) mode in the resonator (inset figure \ref{fig:sim}a), which is used as the heat source. 

The heat source, Q, is defined using equation \ref{ht},

\begin{equation}
    \mathrm{Q} (r,z,t) = {\sqcap}(t) \mathrm{P T_0 F}(r,z),
    \label{ht}
\end{equation}

\noindent
where $\sqcap (t)$ is a step function in time defining the duration for which the laser light is in the microring, P is the absorbed optical power heating the resonator, $\mathrm{T_{0}}$ is the initial temperature of the cavity (room temperature, 298 K), and F(r,z) is the spatial distribution of the optical mode normalized to the total optical energy. The system is allowed to equilibrate with the heat source on until it reaches a steady state, and then the source is turned off and the system temperature is recorded.

The thermal properties of the materials used for main text figure 2e are summarized in table \ref{table1}. 

\begin{table}[h!]
    \centering
    \begin{tabular}{|>{\centering\arraybackslash}p{1.5cm}|>{\centering\arraybackslash}p{1.75cm}|>{\centering\arraybackslash}p{3.5cm}|>{\centering\arraybackslash}p{3.5cm}|}
    \hline
\centering
Material & Density ($\rho$) (kg/m$^3$) & Thermal conductivity (k) (W/m/K) & Specific heat capacity (C) (J/kg/K)\\
\hline

Si$_{3}$N$_{4}$ & 3.29e3 & 2.0 & 700\\
\ce{SiO_2} & 2.2e3  & 1.38  & 703\\
Si &2.3e3 & 130 & 700 \\
\hline
    \end{tabular}
    \caption{Material parameters used in the FEM simulation in the main text.}
    \label{table1}
\end{table}

\noindent
We chose these material properties based on the related literature. For silica and silicon, we used the properties from reference \citenum{huang_thermorefractive_2019} and \citenum{pfeiffer_ultra-smooth_2018} respectively, and we observe that there is little variation in these values within the literature. However, a large variation in the properties of silicon nitride has been reported, possibly due to the impact of deposition process details and conditions. To better understand how the variation in the literature impacts the simulated thermal timescales, we present a comparison between simulation results using material parameters from reference \citenum{huang_thermorefractive_2019}---namely, k$_1$ = 30 W/m/K, $\rho_1$ = 3290 kg/m$^3$, and C$_1$ = 800 J/kg/K, called set P$_1$---and parameters from references \citenum{suyuan_bai_thermal_2009} and \citenum{Drake2020}---k$_2$ = 1.24 W/m/K, $\rho_2$ = 2600 kg/m$^3$ and C$_2$ = 650 J/kg/K called set P$_2$.

Figure \ref{fig:sim}b shows the average temperature of the core as a function of time after the heat source is turned off at $t$ = 0 for both sets of parameters, P$_1$ (red) and P$_2$ (black). For the same optical power, the average temperature of the core prior to cooling is 0.1 K warmer for P$_1$ than for P$_2$. For both, we see fast thermalization followed by a much slower global cooling. To understand how these different parameters would alter the thermal response we observe experimentally, we focus on the fractional change in the first 50 $\upmu$s, seen in the inset of figure \ref{fig:sim}b. Comparing the temperature profile snapshots \ref{fig:sim}c and \ref{fig:sim}d, we see that at $t$ = 0, the core is nearly a constant temperature for P$_1$, as opposed to the gradient present in P$_2$. In both cases however, by 1 $\upmu$s, the hottest part of the ring has shifted into the cladding between the silicon nitride and air. By 3 $\upmu$s, the heat is concentrated at the cladding-air boundary. By 100 $\upmu$s, the system is showing evidence of cooling as a whole rather than thermalization of independent spatial domains.

\section{Soliton lifetime statistics} 

To analyze the effect of mode crossings on soliton lifetime, we record a one second trace in which we sweep the pump laser frequency 660 times, repeated for every heater power. The trace is then divided into the 660 individual sweeps. From there, we identify where the soliton is located in time from the peaks in comb power that correspond to the laser passing through the resonance, and we record $\tau_s$ as described in the main text. The mean comb power during the soliton ``step" is also reported. The number of pulses in the cavity are determined by the comb power. The lowest power and longest duration traces in the data set all have approximately the same comb power. We observe shorter duration states at integer multiples of this comb power, indicating the presence of multiple pulses in the cavity. 

\begin{figure}[h!]
    \centering
    \includegraphics[width=1\linewidth]{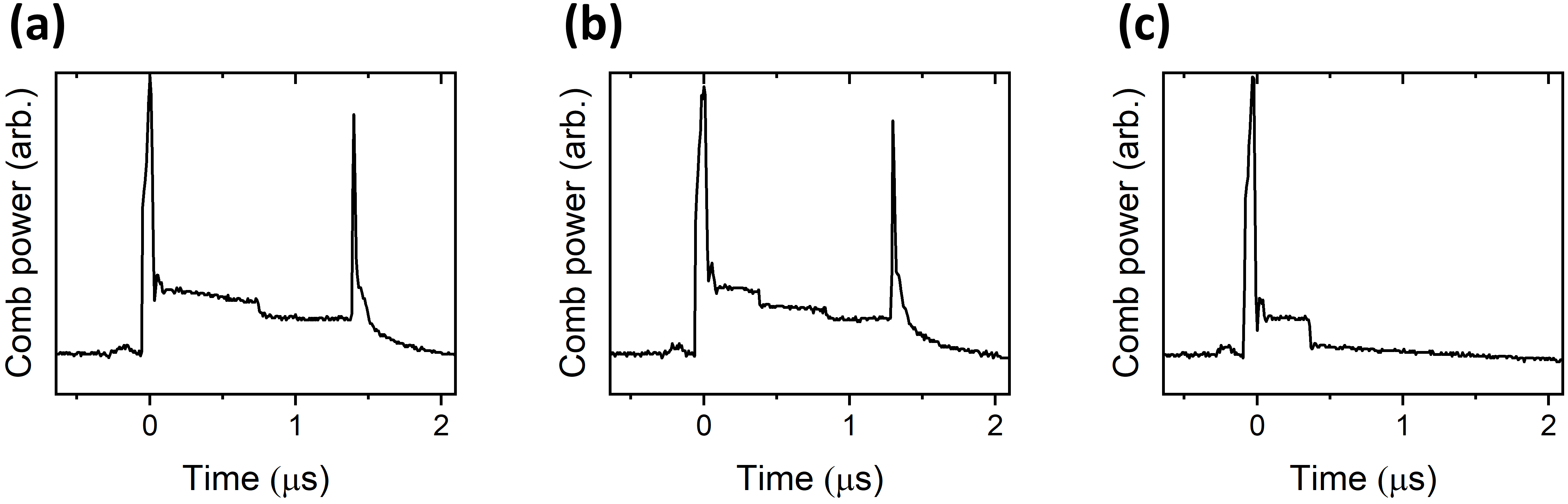}
    \caption{Variety of Soliton States. (a) Short lived soliton state seen in comb power. We see a drop in comb power within the soliton state, indicating a reduction in the number of pulses in the cavity. (b) Another short lived state, with 2 reductions in the number of pulses in the cavity. (c) Soliton states ending with large red pump - resonance detuning.}
    \label{fig:misc_steps}
\end{figure}

Figure \ref{fig:misc_steps} showcases the variety of states we observe. In figure \ref{fig:misc_steps}(a-b) we observe a fast reduction in the number of pulses circulating in the cavity. In figure \ref{fig:misc_steps}c, we see a different kind of step, where we fall out of the soliton state without passing back through the mode. This indicates that the pump is too far red detuned from the mode \cite{Briles2020}. Because we are using thermal recoil as an indication of cooling, if we do not observe the thermal recoil, we exclude them from our analysis. 
\begin{figure} 
    \centering
    \includegraphics[width=1\linewidth]{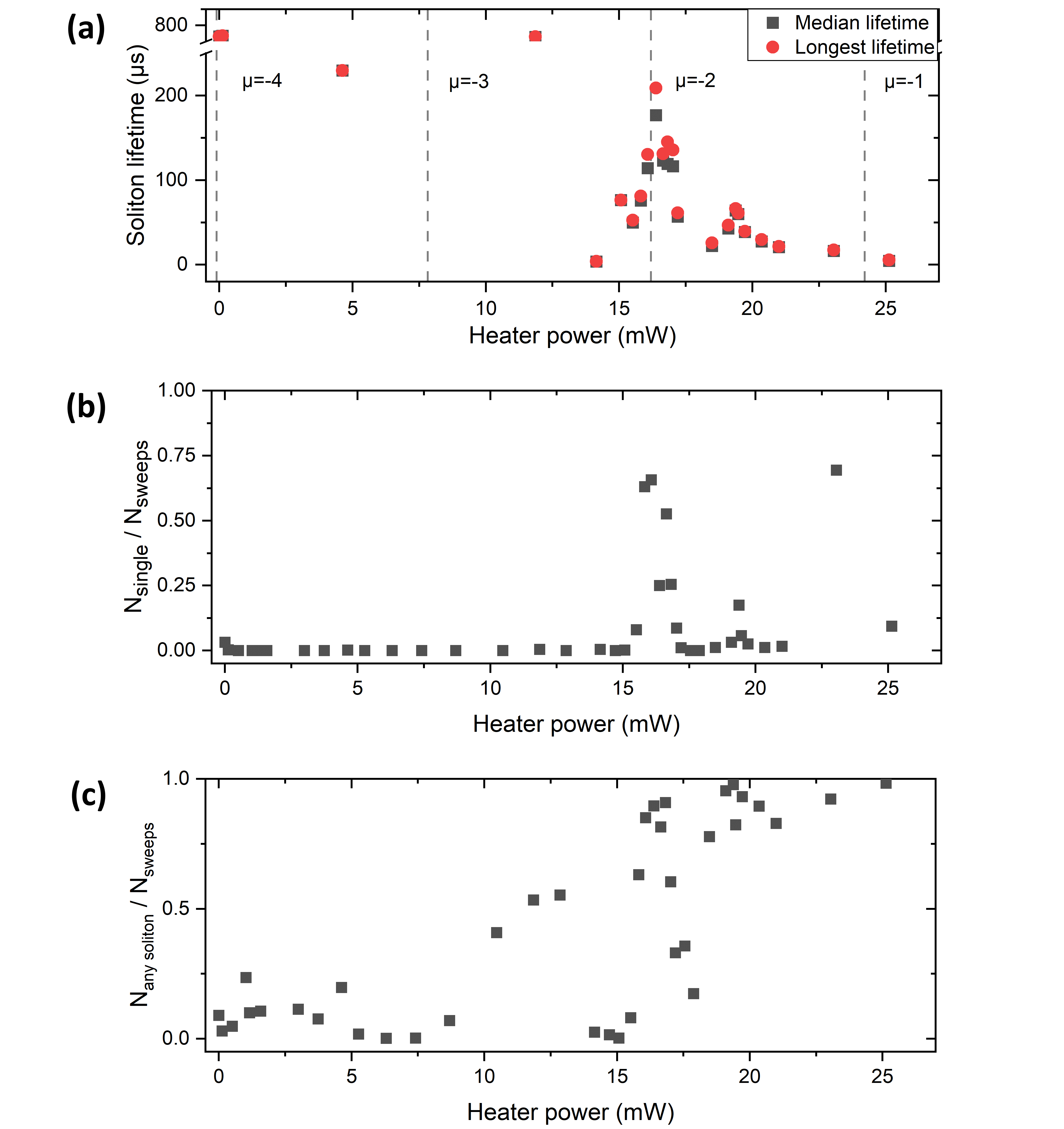}
    \caption{Single soliton lifetime and repeatability with mode crossings (a) Soliton lifetime as a  function of heater power, filtering to include only single soliton states. (b) The fraction of sweeps that resulted in single soliton states. (c) The fraction of sweeps resulting in any soliton state.
    }
    \label{fig:stats}
\end{figure}

Figure \ref{fig:stats}a is a filtered version of main text figure 4a, containing only single soliton steps. We see that a significant number of data points do not contain any single soliton states. Before nearing the $\upmu$ = -2 crossing, single soliton steps are observed rarely. Near the $\upmu$ = -2 and -1 mode crossings, we observe a peak form both in the soliton lifetime and in the repeatability of single soliton states, as seen in figure \ref{fig:stats}b. Probability of creating any soliton state (i.e. with any number of pulses in the resonator) is shown in figure \ref{fig:stats}c.
% we also see in figure \ref{fig:stats}c that the mode crossings have an impact on the repeatability of observing any solitons in the resonator, ranging from nearly 0.1\% to 98.3\%. %[\textcolor{red}{can i claim this about crossings in general since for ALL solitons  there are less clear peaks in repeatability?}] - I'm pretty sure the answer is no. - BS

\newpage
%\section{References}
\typeout{}
\bibliographystyle{apsrev4-1}
\bibliography{supplementalbib}